\documentclass{ifacconf}

\usepackage{amsmath,bm} 
\usepackage{amssymb}  
\usepackage{amsfonts}

\usepackage{graphicx} 
\usepackage{algorithm}
\usepackage{algpseudocode}
\usepackage{etoolbox}
\usepackage{units}

\usepackage{cite}
\usepackage{lpic}
\usepackage{overpic}
\usepackage{mathrsfs}
\usepackage{todonotes}

\usepackage{natbib}        

\providecommand{\eq}{\overset{\Delta}{=}}
\usepackage{bbm}

\usepackage{xcolor}

\allowdisplaybreaks[4]

\DeclareFontFamily{U}{mathx}{\hyphenchar\font45}
\DeclareFontShape{U}{mathx}{m}{n}{
      <5> <6> <7> <8> <9> <10> gen * mathx
      <10.95> mathx10 <12> <14.4> <17.28> <20.74> <24.88> mathx12
      }{}
\DeclareSymbolFont{mathx}{U}{mathx}{m}{n}
\DeclareFontSubstitution{U}{mathx}{m}{n}
\DeclareMathSymbol{\temp}{\mathbin}{mathx}{'341}

{\theoremheaderfont{\upshape} \newtheorem{asmptn}{Assumption}}

\providecommand{\HH}{\mathbb H}
\providecommand{\II}{\mathbb I} 
\providecommand{\PP}{\mathbb P}
\providecommand{\RR}{\mathbb R}

\providecommand{\PP}{\mathbb P}
\providecommand{\RR}{\mathbb R}

\newcommand{\mc}[1]{\mathcal{#1}}
\providecommand{\E}{\mathcal E}
\providecommand{\A}{\mathcal A}

\providecommand{\C}{\mathcal C}
\providecommand{\D}{\mathcal D}

\providecommand{\G}{\mathcal G}

\providecommand{\I}{\mathcal I}

\providecommand{\Pset}{\mathcal P}

\providecommand{\X}{\mathcal X}

\providecommand{\Z}{\mathcal Z}
\providecommand{\U}{\mathcal U}

\newcommand{\Me}[1]{\begin{bmatrix}#1\end{bmatrix}} 

\begin{document}
\begin{frontmatter}
\title{A Framework for Quasi Time-Optimal Nonlinear Model Predictive Control with Soft Constraints} 
%
\author[First]{Joe Ismail}\textsuperscript{,\dag}
\author[First]{Steven Liu} 
\address[First]{Institute of Control Systems, University of Kaiserslautern
P.O. Box 3049, 67653 Kaiserslautern, Germany\\ (e-mail: \textsuperscript{\dag}ismail@eit.uni-kl.de, sliu@eit.uni-kl.de).}
\begin{abstract}                
In many mechatronic applications, controller input costs are negligible and time optimality is of great importance to maximize the productivity by executing fast positioning maneuvers. As a result, the obtained control input has mostly a bang-bang nature, which excite undesired mechanical vibrations, especially in systems with flexible structures. This paper tackles the time-optimal control problem and proposes a novel approach, which explicitly addresses the vibrational behavior in the context of the receding horizon technique. Such technique is a key feature, especially for systems with a time-varying vibrational behavior. In the context of model predictive control (MPC), vibrational behavior is predicted and coped in a soft-constrained formulation, which penalize any violation of undesired vibrations. This formulation enlarges the feasibility on a wide operating range in comparison with a hard-constrained formulation. The closed-loop performance of this approach is demonstrated on a numerical example of stacker crane with high degree of flexibility.
\end{abstract}
\end{frontmatter}
%
%
%
%
\section{INTRODUCTION}
%
%
%
%
%
Ever since the \emph{Acta Eruditorum} of 1697, where Johann Bernoulli posed the brachistochrone problem with minimum-time trajectories, and thus the rise of calculus of variations, time-optimal control was of a great interest. Well-known solutions established by Pontryagyn's maximum principle provides sufficient and necessary conditions for the bang-bang property of the optimal control inputs. 
Classical solution techniques are based on indirect methods with explicit control law. For nonlinear systems, numerical solutions became inevitable. The first extensions to nonlinear model predictive control (NMPC) distinguished between the direct and indirect methods to obtain a numerical solution of the optimal control problem (OCP). While indirect methods encompass also the well known calculus of variations, direct methods transform the OCP through discretization into a finite dimensional nonlinear program (NLP), \cite{diehl2017numerical}. This paper focuses on time-optimal control of discrete nonlinear dynamic systems using direct methods.
Some previous studies dealt with time-optimal control problems (time-OCP) in the context of model predictive control (MPC). \cite{van2011model} introduced in his approach a two-layer time-optimal MPC for linear systems, while the upper layer determines the optimal horizon length, the lower layer contains a tracking MPC. Beside dealing with linear system, the changing of horizon length from one problem to another is a draw back of this approach. \cite{zhao2004nonlinear} introduced a quasi time-optimal NMPC, which is formulated in standard regulator NMPC form, with time-dependency on the terminal cost and fixed horizon length. \cite{verschueren2016time2} computed time-optimal motions along a Cartesian path for robotic manipulators and in \cite{verschueren2016time1} the time-OCP was stated in a path-parametric formulation to compute motions around a periodic optimal trajectory for a race car. \cite{lipp2014minimum} focused on a subset of trajectory generation problems in which a specific path is already given. \cite{rosmann2017time} used a dynamic grid time-optimal control formulation to reduce the number of control interventions. Especially in the domain of vibration control, recently NMPC gained an increasing attention to replace the widely established input shaping. Input shaping is generated for a specific frequency and damping ratio and cannot deal with time-varying frequencies. Among the previous mentioned studies, \cite{zhao2004nonlinear} mentioned the problem of vibrations arising with time-optimal NMPC approaches. \cite{lau2003input} gave analytical solution of a linear system showing a zero residual vibration behaviors of open-loop trajectories of time-OCP. However, even open-loop trajectories might excite undesired resonance frequencies or residual vibrations for both weakly damped flexible systems or unmodeled mode vibrations. For such cases, no vibrational guarantees are studied so far.   
The present paper presents a novel time-optimal NMPC approach for point-to-point transitions of nonlinear dynamic systems in a soft-constrained NMPC framework. This approach addresses the issue of vibrations, that arise especially for bodies with high degree of flexibility.
The application of this paper deals with stacker cranes (STCs), which are tasked to maximize the productivity by executing fast and accurate positioning maneuvers. Additionally promoted from the mast flexible and slender form of the beam, time-optimal control with bang-bang nature may, however, might excite undesired residual mode vibrations or excite undesired resonance frequencies. 
Furthermore, STCs have a time-varying vibrational behavior due to the variable lift load in position and weight. The model used here captures the steadily changing vibrational behavior of the STC and has nonlinear dynamics. This paper extends the frame of the active vibration damping control conducted using a soft-constrained NMPC, proposed in \cite{ismail2019control}, however, with time-optimal objectives. Time-optimal control approaches are important objectives for cranes, for which a vast number of studies exist. Within MPC, only few studies have been conducted, like \citep[][]{van2011model, van2011experimental, vukov2012experimental, kapernick2013model}. 
%
%
%
%
%
%
%
%
%
%
%

\textit{Contribution:}
%
%
%
%
%
%
%
The main contribution of this paper is the introduction of soft-constrained time-optimal NMPC. In this configuration, undesired vibrations, typically resulting from the aggressive optimal control input are addressed explicitly and considered with soft-constraints. 
Dealing with undesired vibrations as soft-constraints enlarge the feasibility domain, especially for mechanical systems with low resonance frequencies. The proposed approach is applied on two types of time-optimal control problems. To the best of author's knowledge, this study is the first attempt to present a soft-constrained time-optimal NMPC with vibrational behavior consideration.  
%
%
%
The rest of the paper is structured as follows: Section \ref{sec:Preliminaries} provides some preliminaries and introduces some ingredients, which are used in Section \ref{sec:TOMPC}. In Section \ref{sec:TOMPC} the proposed approach is detailed. Section \ref{sec:result} and Section \ref{sec:result2} shows the simulation results on two numerical problems. Finally, Section \ref{sec:conc} provides a conclusion.
%
%
%
%

%
\section{Preliminaries}
\label{sec:Preliminaries}
\subsection{Mathematical Notations}
\label{subsec:notations}
%
%
Let $\II$ denote the set of non-negative integers. For $n, m \in \II \cup \{\infty\}$, let $\II_{\geq 0}$ and $\II_{n:m}$ denote the sets $\{ r \in \II: n \leq r \leq m\}$, respectively. Similarly, $\RR_{\geq 0}$ denote the non-negative real numbers, $\RR^{n}$ are real valued $n$-vectors, $\PP_N(x, r)$ is an MPC optimization problem with horizon $N$, initial state $x$ and reference $r$. $||.||$ and $\| . \|_\infty$ are the $l_2$ and the $l_\infty$ norms. A set $\A \subset \RR^n$ is a C-set, if it is convex and contains the origin. A convex $\HH$-polytope denotes a bounded intersection of $q$ closed half-spaces $\Pset=\{x \in \RR^n: Cx\leq d, C\in \RR^{q \times n}, d \in \RR^q\}$. $\otimes$ is the Kronecker product. $\I$ is the unity matrix.  
%
%
%
%
%
%
\subsection{Problem Statement}
\label{sec:ProblemFormulation}
%
%
%
%
Now consider a discrete time system dynamic, described as,
\begin{align}
\label{eq:ODEMS}
	x_{k+1} &= f(x_k, u_k),~~\text{with}~~ x(0)=x_0, 
\end{align}
where $f(.):\RR^n\times\RR^m \rightarrow \RR^n$ and $f(0,0)=0$, with $k \in \II_{\geq 0}$ as the discrete time index, $x_k \in \X \subset \RR^n$ is the state variable, $u_k \in \U \subset \RR^m$ is the control input. $\X,~\U$ are state-, input sets, respectively, which are assumed to be known exactly and are PC-polytopic sets, respectively. $x_k := \phi(k;x_0,\bm u)$ is the solution of \eqref{eq:ODEMS} at time $k$ with initial state $x_0$ and input sequence $\bm u$.
More general hard mixed constraints $\Z \subseteq \X \times \U$ are imposed on the state-input space, given in polytopic representation,
\begin{align}
(x_k,u_k) \in \Z := \C_x x_k + \D_u u_k \leq \E ,~k \in \II_{\geq 0}, \label{eq:MixedConstraints}
\end{align}
with $\C_x \in \RR^{n_e\times n}, \D_u \in \RR^{n_e\times m}$ and $\E\in \RR^{n_e}$ to be the matrix element of the polytope in state-input space given in $\HH$-representation. $N$ is a given integer i.e. horizon. Additionally, a polytopic terminal constraint set is introduced as,
\begin{align}
\X_f:= \G_x x_k \leq \mc F,~k \in \II_{N:\infty}, \label{eq:TerminalConstraints}
\end{align}
with $\G_x \in \RR^{n_f\times n}, \mc F \in \RR^{n_f}$, which specifies the hyperplanes bounding of $x_N \in \X_f \subset \X$.
The purpose of the controller is to smoothly steer a system from some initial state $x_0$ to a neighborhood of a terminal set $\X_f \in \X$ within a minimal transition time $T$ and with the minimal possible vibrations, while satisfying the given input and state constraints. For point-to-point motions $\X_f \in {x_f}$ with $x_f \in \X$ as the terminal state. The controller objective of fast maneuver transitions and minimal vibrational damping has an antagonistic nature. 
%
%
%
%
\subsection{Vibrational Frequencies Prediction}
\label{subsec:EPSDP} 
%
%
The idea of vibration damping control based on predictive control leads naturally to the idea of vibrational prediction, which is incorporated in the NMPC in Section \ref{sec:SoftMPC}. Due to parameter variations e.g. position of a crane load, lot of systems experience time-varying vibrational behavior and thus time-varying resonance frequencies. For this, parameter variation $\varrho$, are considered and \eqref{eq:ODEMS} can be rewritten as,
\begin{align}
\label{eq:ODEMS2}
x_{k+1} &= f(x_k, u_k, \varrho_k),~~\text{with}~~ x(0)=x_0. 
\end{align}
Starting from known initial value $x_0$ and parameter set $\varrho_0$ and for a given input sequence, the differential equation is solved numerically and the states prediction propagate as $x_k := \phi(k;x_0, \varrho_0, \bm u), k \in \mathbb{I}_{0:N}$. For this, the well-known multiple shooting with Runge-Kutta integrator is used over a fixed horizon length with $N$ intervals. Based on the state predictions, it is possible to predict the vibrational behavior and the associated frequencies. Vibrational prediction is carried out in this paper in terms of power spectral density (PSD). The main idea of spectral analysis is to decompose a signal into a sum of weighted signals, which allow to access each frequency content separately. One of the most commonly used method for estimating the power of a signal is the Welch's method, sometimes also referred as the periodogram averaging method. Due to the sinusoidal nature of vibrations a sinusoidal fitting function is applied before conduction PSD, which enhance the signal quality and thus the vibrational frequency prediction. Its worth to mention, that frequencies prediction can be improved using wavelet transformation, since this gives more insights about each frequency characteristic within a short prediction.    
\begin{asmptn}
For a finite horizon problem, it is assumed that the optimization horizon $N \in \mathbb I_{\geq 0}$ is large enough to conduct a good vibrational frequency prediction performance and also to enforce convergence.
\end{asmptn}
%
%
%
%
%

%
%
\section{Time-Optimal Model Predictive Control}
\label{sec:TOMPC}
In this Section, the control problem, proposed in Section \ref{sec:ProblemFormulation} as a time-optimal control problem (OCP) is tackled. It is well-known, that time-OCPs are classically assigned to indirect methods, by which variational problems are reduced i.e. according to Pontryagin maximum principle into a boundary value problem (BVP) with initial conditions on the state and final conditions on the co-state. The resulting BVPs are difficult and thus commonly solved numerically based on solving sequentially many forward and backward initial value problems (IVP), and then plugging their solutions together. Additional limitation of indirect methods is the treatment of mixed path constraints, \cite{diehl2017numerical}. For nonlinear systems, obtaining the necessary optimality conditions and thus a time-optimal control law in closed form is indeed more challenging, which is another drawback of indirect methods. Despite the approaches, which use indirect methods for nonlinear MPC implementations, like in \cite{kapernick2014gradient}, indirect methods still require a priori computations and are therefore still less attractive for MPC.
In contrast, the efficient direct multiple shooting (MS) has become popular as a highly competitive approach. MS provides an approximated solution of the BVP. For this reason, MS is applied in this paper to solve the time-OCP. 
%
%
%
\subsection{Time-OCP Formulation}
\label{sec:subTOMPC}
%
%
%
In time-OCPs, it is required to minimize the travel time $t \in \mathbb R_{0:T}$ over a transition time $T$, which a system requires to move from some initial state $x_0$ to a neighborhood of some terminal state $x_f \in \X_f$. This implies that the total transition time $T$ is considered as a decision variable, such as,
\begin{align} 
\underset{\bm{x}, \bm{u}, T}{\text{min}}~\int^{T}_{0} 1 \mathrm{d}t \eq \underset{\bm{x}, \bm{u}, T}{\text{min}}~T. \label{eq:TOCPCost}
\end{align} 
Using MS, the original infinite time-OCP is transformed into a finite dimensional nonlinear programming problem (NLP). On the one hand, the time horizon is divided into $N$ shooting intervals $[t_k,t_{k+1}], k \in \II_{0,N-1}$. On the other hand, the control input trajectory is parametrized by assuming it to be a piece-wise constant signal over each interval. Each interval induces an IVP with an artificial initial value, which can be solved by well-known integrators, e.g., Runge-Kutta or explicit Euler. To ensure compliance of continuity between subsequent intervals, additional matching conditions are obtained which incorporate in the NLP as equality conditions. These matching conditions are shooting nodes and are used in this paper. For the sake of simplicity, however, these additional conditions are not explicitly mentioned in the remainder of this paper. Interested reader is refereed to \cite{diehl2017numerical}. In this paper, solutions with a receding horizon fashion are targeted by repeatedly solving the NLP numerically. The resulting time-OCP structure is given as,
\begin{subequations}
\label{eq:TOCP1}
\begin{align} 
\PP_N(x_0,.)&: \mathcal{J}^*_N =~ \underset{\bm{x}, \bm{u}, T}{\text{min}}~T \label{eq:TOCP2} \\ 
\text{s.t.} &~~x(0)= x_0,  \label{eq:TOCP3} \\
&~~x_{k+1}=f(x_k, u_k), \label{eq:TOCP4}  \\
&~~x_{N} \in \mathcal{X}_f \subseteq \X, \label{eq:TOCP5} \\
&~~(\bm{x}, \bm u)\overset{\Delta}{=} (x_k, x_{N}, u_k) \in \Z, \label{eq:TOCP6}
\end{align}
\end{subequations}
$\forall k \in \mathbb I_{0:N-1}$. The solution of \eqref{eq:TOCP1} is due to nonlinearity certainly not unique, but the control is often still found to have a bang-bang nature.
%
%
\subsection{Quasi Time-Optimal NMPC}
\label{sec:SoftMPC}
%
%
%
Linear time-scaling is an important technique to solve time-OCPs, by which the free final time is transformed to a fixed final time. This technique work around the fact that the transition time $T$, which is considered as decision variable is a priori not known. In addition, scaled time provide an important property, by which the underlying shooting interval size is independent of $T$ and thus fixed. Accordingly, both time variable and system dynamics are scaled linearly by the optimization variable $T$. This transformation is given as,
\begin{align} 
t(\tau) :=\tau T,~~~~\forall \tau \in \mathbb R_{0:1},
\label{eq:Timetransformation}
\end{align}  
with $\tau$ as the new pseudo-time variable. Accordingly, the state and control become $x(t(\tau)):= x(\tau T)$ and $u(t(\tau)):= u(\tau T)$. Time-optimal objective become $\underset{T}{\text{min}}~\int^{1}_{0} T \mathrm{d}\tau$. Subsequently, corresponding ODE is transformed to,
\begin{align} 
 \frac{\mathrm{d} x(\tau)}{\mathrm{d} \tau} &= f(x(t),u(t))T,~~~~~~~~~\forall t \in \mathbb R_{0:T}, \nonumber \\ 
& \eq f_s(x(\tau), u(\tau), T),~~~~\forall \tau \in \mathbb R_{0:1}. 
\label{eq:scaleddynamic}
\end{align}  
The time variable $t \in \mathbb R_{0:T}$ with N shooting intervals is mapped on $\mathbb R_{0:1}$ with the same number of shooting intervals, however, each with a size of $1/N$. Consequently, numerical integration is conducted based on the new variable $\tau$, which is independent of decision variable $T$. 
One idea to express the whole transition time $T$ is given in \cite{verschueren2018convex}, by which a $T$ is separated into a sequence of decision variables $T_k$, each defined over a shooting interval. The reason for this, is to preserve the shooting structure and to guarantee a fixed sampling time.
In the context of NMPC, and in order to consider the vibrational behavior and try to avoid the undesired ones, one might consider undesired vibrations, like resonance frequencies as hard constraints. However, especially mechanical systems suffer mostly from comparable low vibrational and thus low resonance frequencies. Therefore, such consideration shrinks the operational domain. This can be avoided by softening these constraints. Thus, the resonance frequencies defined in \eqref{eq:PolyApprox} are considered as inequality constraints with infeasible boundaries and relaxed by introduction of slack variables $s_k, \forall k \in \mathbb I_{0:N}$. The purpose of relaxing these inequality is twofold: first one, is to ensure feasibility of the MPC problem in a large region of the state space by tolerating temporary constraint violations; the second, is to mitigate the impact of the bang-bang control appears in context of time-OCP on excitation of the undesired frequencies like resonance. Additionally, and according to the technique given in Section \ref{subsec:EPSDP}, a vibrational frequency prediction is repetitively conducted over each horizon to estimate the future frequencies. To reduce any constraints violation, the slack variables are penalized and included in the cost function. 
The proposed soft-constrained NMPC problem denoted as $\mathbb {P}_N(x_0, .)$ is formulated as:     
%
%
\begin{subequations}
\label{eq:NLPEquation}
\begin{align} 
\mathcal{J}^s_N &\overset{\Delta}{=}~ \| s_N \|^2_P +\sum_{k=0}^{N-1} \frac{T_k}{N} + \| s_k \|^2_{\frac{1}{N} \cdot S} \nonumber \\
\PP_N(x_0,.): ~\mathcal{J}^*_N &= \underset{\bm{u},\bm{x},\bm{s}, \bm{T}}{\text{min}}~  \mathcal{J}^s_N(\bm{u},\bm{x},\bm{s}, \bm{T}) \label{eq:NLPEquation1} \\
\text{s.t.} &~~x_0=x,  \label{eq:NLPEquation2} \\
&~~x_{k+1}=f_{s}(x_{k},u_{k},\varrho_k, T_k), \label{eq:NLPEquation3}  \\
&~~ \mathcal{E}(x_k,u_k,\varrho_k)-\mathcal B^p_- \geq -s_k, \label{eq:NLPEquation4}  \\
&~~ \mathcal{E}(x_k,u_k,\varrho_k)-\mathcal B^p_+ \leq s_k, \label{eq:NLPEquation5} \\
%
&~~ \mathcal{Z}:=\C_x x_k + \D_u u_k \leq \E, \label{eq:NLPEquation5_1}  \\
&~~(\bm{x}, \bm{u})\overset{\Delta}{=} (x_k, x_{N}, u_k) \in \mathcal{Z}, \label{eq:NLPEquation5_2} \\
&~~(s_k,s_N, T_k) \geq 0, \label{eq:NLPEquation6} \\
&~~x_{N} \in \mathcal{X}_f \subseteq \X,
\end{align}
\end{subequations}
$\forall k \in \mathbb I_{0:N-1}$, where $\bm{x} \in \mathbb R^n$, $\bm{u} \in \mathbb R^m$, $\bm{s} \in \mathbb R^s$ and $ \bm{T} \in \mathbb R$. Note that, $f_{s}$ is the discrete time representation of the time-scaled dynamic from \eqref{eq:scaleddynamic}. Weighting are chosen as, $S\in \mathbb R^{s\times s}, S\succeq 0 $ for the stage cost and $P\in \mathbb R^{s\times s}, P\succeq 0 $ for the terminal cost. Additionally, $\Z \subseteq \X \times \U$ denotes the hard mixed constraints for admissible state and control sets in the state-input space. $\mathcal{X}_f \subseteq \mathcal{X}$, defined as in \eqref{eq:TerminalConstraints}, is a compact terminal set computed as a positive control invariant set using the solution of the discrete Riccati equation of the linearized system dynamics in the vicinity of the terminal state $x_f$.  
Furthermore, $\mathcal{E}(x_k,u_k, \varrho_k)$ states the solution of the frequency prediction from Section \ref{subsec:EPSDP} and form together with the boundaries ($\mathcal B^p_-, \mathcal B^p_+$) of undesired frequencies in \eqref{eq:NLPEquation4} and \eqref{eq:NLPEquation5} the soft inequality constraints. It is worth to mention that, the underlying model for vibrational frequency prediction uses beside open-loop state and input the variable $\varrho_k$ to conduct an IVP based on the non-scaled time $t \in \mathbb R_{0:T}$ to carry out a proper frequency prediction. The inequalities describe the soft-constrained space, which can be time-variant as in Section \ref{sec:NaturalFrequencies}. 
Problem $\mathbb {P}_N(x_0, .)$ induce implicitly a set of feasible state, slack and time sequences $\bm{x}^*:\mathbb R^n \mapsto \mathbb R^{N\times n}$, $\bm{s}^*:\mathbb R^s \mapsto \mathbb R^{N\times s}$, $\bm{T}^*:\mathbb R \mapsto \mathbb R^{N}$ respectively. Similarly, a set of feasible control sequence $\bm{u}^*=\{u^*_0, u^*_1,\dots,u^*_{N-1}\}$, with $\bm{u}^*:\mathbb R^m \mapsto \mathbb R^{N \times m}$ is induced. Only the first control entry of each induced control sequences is applied to the plant as a closed-loop control. After that, the optimization horizon is shifted toward the next discrete time-instant in a receding horizon fashion. However, to ensure a compliance with the sampling-time $T_s$ from the plant the time is rescaled by $N\cdot T_s$
first $T_0$ must be constraint to be equal $N\cdot T_s$, which is a rescale, while the remaining $T_k$ are free, however equidistant. Additional conditions to sustain equidistant shooting intervals is not strictly necessary, whoever such a condition prevents the solver from exploiting the time scaling to creating shortcuts through constraints.  
Note that, in case of plant-model mismatches, tracking control with a simple and computational cheap state-feedback optimal controller gain $K$ of the discrete Riccati equation would be necessary using the dual-mode paradigm, i.e. Mode 1 ($u_k=Kx_k,~\forall  k \in \mathbb I_{0:N-1}$). After reaching the neighborhood of terminal state $x_f$, Mode 2 ($u_k=K(x_k-x_k^*)+u_k^*,~\forall  k \in \mathbb I_{\geq N}$) is activated.
%
%
%
\subsection{Quasi Time-Optimal NMPC with Free Final Time}
\label{sec:QuasiSoftMPC}
%
%
For the sake of completeness, the idea of the soft-constrained NMPC from Section \ref{sec:SoftMPC} is carried out on the mostly used quasi time-optimal control with free final time $T$. Quasi time-optimal control contains a standard running cost function with a quadratic form and free terminal time in the terminal cost. With appropriate weighting, time optimality can dominate at the beginning of the trajectory. Nevertheless, slightly modifications of the objective are carried out here. This formulation addresses also mitigation of bang-bang control and thus limiting the excitation of undesired vibrations. In contrast to the formulation in the previous Section, this formulation does not employ a time scaling and uses instead the discrete-time dynamics . In the same manner as quasi time-optimal control, the decision variable $T$ is considered as a dummy state, which is introduced in the terminal cost. Furthermore, the usual class of quadratic objective is utilized for the stage cost with the soft-constrained extension. The proposed soft constrained NMPC problem denoted as $\mathbb {P}_N(x_0, .)$ is similar to $\mathbb {P}^Q_N(x_0, .)$ in Section \ref{sec:SoftMPC} and formulated as:     
%
%
%
\begin{subequations}
\label{eq:QuasiTimeOCP}
\begin{align} 
\PP^Q_N(x_0,.):  \mathcal{J}^s_N \overset{\Delta}{=}  \mathcal{J}_N &~(x_{T}, s_{T}) +\sum_{k=0}^{N-1} l_k(x_k, u_k, s_k) \nonumber \\
\mathcal{J}^{q*}_N = \underset{\bm{u},\bm{x},\bm{s}, T}{\text{min}}&~ \mathcal{J}^s_N(\bm{u},\bm{x},\bm{s}, T) \label{eq:QuasiTimeOCP1} \\
\text{s.t.} &~~x_0=x,  \label{eq:QuasiTimeOCP2} \\
&~~x_{k+1}=f(x_{k},u_{k},\varrho_k), \label{eq:QuasiTimeOCP3}  \\
&~~\eqref{eq:NLPEquation4}-\eqref{eq:NLPEquation5}, \label{eq:QuasiTimeOCP4}  \\
&~~(s_k,s_T, T) \geq 0, \label{eq:NLPEquation6}   \\
&~~x_{T} \in \mathcal{X}_f \subseteq \X,
\end{align}
\end{subequations}
$\forall k \in \mathbb I_{0:N-1}$, with $\mathcal{J}_N \overset{\Delta}{=} T_F + \| x_N-x_f \|^2_Q + \| s_N \|^2_P$, $ l_k \overset{\Delta}{=}  \| x_k-x_s \|^2_Q + \|  u_k-u_s \|^2_R + \| s_k \|^2_S$. To guarantee convergence with a finite prediction horizon, it is required that the state reaches the desired endpoint at the end of the prediction horizon. Compared to Section \ref{sec:SoftMPC}, additional weightings are introduced, which are chosen as, $Q\in \mathbb R^{n\times n}, Q\succeq 0 $, $R\in \mathbb R^{m\times m}, R\succ 0 $ for weighting the stage costs. Additionally, $P\in \mathbb R^{n\times n}, P\succeq 0$ and a scalar $F$ for weighting the terminal costs. 
%
%
%
%
%
\subsection{Stability and Recursive Feasibility}
\label{sec:StabilityandSoon}
%
%
Recursive Feasibility is a necessary property to mimic the infinite horizon MPC. However, time-optimal approaches cannot ensure recursive feasibility, since the length of shooting intervals decrease while the overall transition time decrease during the closed-loop control, which is also the case for time-transformation method. Fixing the sampling time is more realistic, but it might even increase this effect. Additional measurements of introduction of time-varying terminal invariant sets or introduction of controllable invariant sets, similar to the tubes used in robust MPC, solve this problem. Furthermore, recursive feasibility cannot be guaranteed when the closed-loop control is in the vicinity of the optimal transition, as a recent study \cite{rsmann2020stabilizing} shows. However, in practice with both, the introduction of $\X_f$ as a time-invariant terminal set and fine shooting intervals, where optimality often holds in an approximative manner recursive feasibility might be implied. Time-invariant terminal set are conservative, since the time-optimal problem is inherently time-varying. Due to the difficulty in establishing of recursive feasibility, no stability proofs exist so far. On the other side, quasi time-optimal NMPC with free final time and conventional quadratic cost function retains the properties used for classical MPC, which makes recursive feasibility and stability proofs easier to establish.
%
%
%
%
%
\section{NUMERICAL CASE STUDY: Stacker Crane}
\label{sec:Setup}
%
%
%
In this case study, the proposed soft-constrained time-optimal NMPC is studied on a numerical example of a stacker crane (STC) with high degree of flexibility. Some key ingredients, which are used in the time-optimal NMPC setup, are introduced firstly.
%
%
%
%
\subsection{Dynamic}
\label{subsec:dynamicFormulation}
\begin{figure}[htbp]
\centering
\includegraphics[scale=0.45]{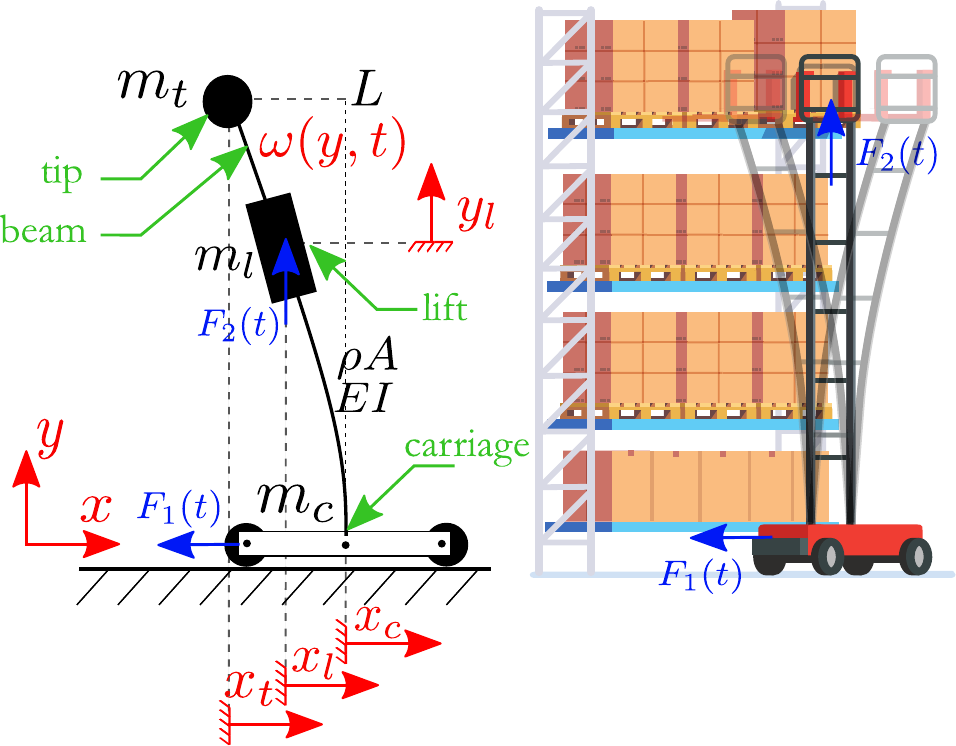} 
\vspace{-2mm}
\caption {flexible stacker crane (right), illustration (left)}
\label{fig:CTS}
\end{figure}

%
%
%
%
STCs are considered mostly as a single Euler-Bernoulli beam with fixed boundary conditions and thus an a priori known vibrational behavior. Moving lift position and load changes lead to variable vibrational behavior. To capture these changes a model of two Euler-Bernoulli beams with time-varying boundary conditions was considered in \cite{ismail2019model} and is adapted here. Fig. \ref{fig:CTS} depicts the model of the STC. Both the carriage and the lift positions are controllable in terms of actuation. The actuators apply axial forces, denoted as $F_1$ and $F_2$. $(x_t,L), (x_l,y_l)$ and $(x_c,0)$ denote the coordinates of the tip, lift and carriage respectively. $EI$ represents the flexural rigidity. Also, denoted are the cross-sectional area of beam $A$ and density $\rho$. For the beam of length $L$, the spatial variable is given as $y \in \mathbb R_{0:L}$. $\omega(y,t)=\omega_y$ denote the absolute axial deflection. Carriage, lift and the tip mass are modeled as lumped masses $m_c$, $m_l$ and $m_t$ respectively. The flexibility of STC is modeled as a distributed parameter system (DPS). The control-oriented model gives a lumped form of the DPS to obtain ordinary differential equations (ODEs). 
Accordingly, the spatial dependency in $\omega(y,t)$ is approximated by a finite degree of freedom of two modes $a_i$ and two shape functions $\psi_i$ as,  
\begin{align}
\label{eq:AxialDeflection}
\omega(y,t) = \sum_{i=1}^2 \psi_i(y) \cdot a_i(t).
\end{align}
With $q(t) = \begin{pmatrix} x_c(t) & y_l(t) & a_1(t)  & a_2(t) \end{pmatrix}^T$ the resulting equations of motion i.e. ODEs are rewritten as,
\begin{align}
M (q) \ddot{q} + C(q,\dot{{q}}) + K q =  F_{ext}, \label{eq:systModell}  
\end{align}
with the inertia matrix $M(q) \in \mathbb R^{4\times 4}: M^T=M$, the coriolis matrix $C(q,\dot{q})  \in \mathbb R^{4} $ and the stiffness matrix $K \in \mathbb R^{4\times 4}: \{K_{ij} \in \mathbb R^{2\times 2} \land K_{ij}^T=K_{ij}=0_{2 \times 2}, \forall i=j: i,j \in \mathbb I_{1:4} \setminus K_{44}\}$. Force is given as, $ F_{ext}=\begin{pmatrix}F_1 & F_2 & 0  & 0\end{pmatrix}^T.$ Details are given in \cite{ismail2019model}.
Equations of motion given in \eqref{eq:systModell} are rearranged as $\ddot{q} = f(q,\dot{q},F_{ext})$, with $\dot{ q}(0)=\dot{q}_0, q(0)=q_0$, then the second order state vector is rewritten in a lighter notation of two first order states, which results an eight-order system, given as,
\begin{align}
\label{eq:ODEMScont}
\dot{x} &= f(x, u),~~\text{with}~~ x(0)= x_0,
\end{align}
%
%
%
%
\subsection{Time-Varying Resonance Frequencies}
\label{sec:NaturalFrequencies}
%
%
%
%
%
Frequency responses and specifically resonance frequencies are sensitive to the operational point of the lift $y_l$ and to its mass $m_l$. $\varrho=[m_l, y_l]^T$ denotes the parametric variation of the mass as well as the operational range of the lift, considered to be over a domain set $\mathcal{D}$ with an upper and lower boundary $(\underline{ \varrho},\overline{\varrho}) \in \mathcal{D} \subseteq \mathbb R^2$. Thus, $\varrho \in [\underline{\varrho},\overline{\varrho}]$, $\underline{\varrho} \in \mathbb R_{\geq 0}$ and is time-varying. In \cite{ismail2019control} the sensitivity of resonance frequency due to the parametric variation was studied. Accordingly, a characteristic equation of a nonlinear algebraic equation (NAE) nature is derived analytically, given as,
\begin{align}
\mathcal G(y_l,m_l,\Omega)=0.
\end{align}
For each fixed pairs $y_l$ and $m_l$, infinite solutions of resonance frequencies $\Omega \in \mathbb R^\infty$ are obtained. These solutions are carried out numerically in terms of an iterative nonlinear root finding. This provides a set of algebraic hypersurfaces $\mathcal S_{s}: s \in \mathbb I_{\geq 0}$, which are fully calculated across the entire variation range of $m_l$ and the operating range of $y_l$. Each of these algebraic hypersurfaces is an algebraic variety contained in affine space as, 
\begin{equation}
\mathcal S:=\{(m_l,y_l,\Omega) \in \mathbb R^3:\forall \mathcal G \in \mathcal S,~\mathcal G(y_l,m_l,\Omega)=0 \}. \label{eq:hypersurfaces}
\end{equation}
Refer to remark \ref{rem:hypersurfaces}, the boundaries ($\mathcal B^p_-, \mathcal B^p_+$) are in the Minkowski spacetime $\mathcal H$, and thus time-variant. 
%
%
\begin{rem}
\label{rem:hypersurfaces}
Recall \eqref{eq:hypersurfaces}, due to the prediction, hypersurfaces $\mathcal S_s,~\mathcal P_p$, and thus the boundaries $\mathcal B^p_-,~\mathcal B^p_+$ become in fact defined in the Minkowski spacetime $\mathcal H \in \mathbb R^{3,1}$, with four-dimensional manifold (three dimensions of space and one dimension of time $t$) $\mathcal S_s, \mathcal P_p, \mathcal B^p_-,\mathcal B^p_+ \in \mathbb R^3 \subset \mathcal H $ with $m_l,y_l, \Omega$ as elements of $\mathbb R^3$. The spacetime is defined as,
\begin{align*}
\mathcal H:=\{ m_l,y_l,\Omega, t \in \mathbb R:~\bigcup~\{\mathcal S_s, \mathcal P_p, \mathcal B^p_-,\mathcal B^p_+ \} \times \{t\} \}.
\end{align*}
\end{rem}
%
%
%
%
\subsection{Inequality Constraints: Over- and Under approximation}
\label{sub:PolyApprox}
\begin{figure}[htbp]
\centering
\includegraphics[scale=0.45]{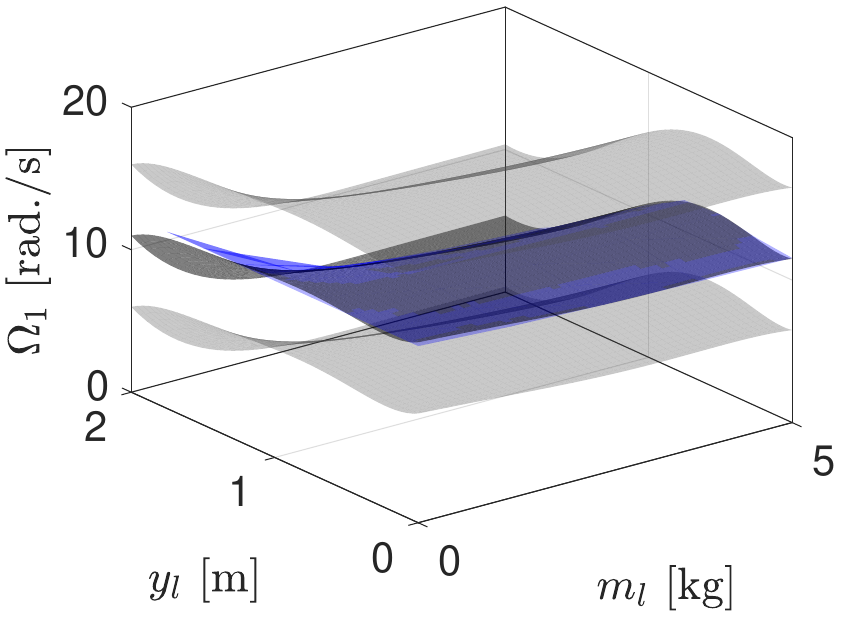} 
\caption {resonance frequency (blue), approximation (black), boundaries (gray)}
\label{fig:FreqBoundaries}
\end{figure}

%
%
The purpose of providing a polynomial approximation is twofold: first one, to replace the computational expansive iterative nonlinear root finding by a polynomial function with a cheap mapping; the second, to define an over and under approximation boundaries of the resonance frequency hypersurfaces. 
These boundaries are coped in the optimization problem in Section \ref{sec:SoftMPC} as infeasible inequality constraints. For the approximation the standard approach of least squares has been carried out, by minimizing $\| \mathcal S_s(\underline{\varrho},\overline{\varrho}) - \mathcal P_p (\underline{\varrho},\overline{\varrho}) \|^2$. As a result, following approximation is obtained,
\begin{align}
\mathcal P_p = \sum_{i, j\in{\mathbb{N}}}^{p} c_{i,j}m_l^i y_l^j.
\end{align}
A satisfactory approximation was achieved by a 5th ($p=5$) order polynomial as Fig. \ref{fig:FreqBoundaries} illustrates. The boundaries are defined as a lift-up and down of the approximation,   
\begin{align}
\label{eq:PolyApprox}
\mathcal B^p := \{ \mathcal B^p_- \leq \mathcal P_p \leq  \mathcal B^p_+: m_l, y_l \in \mathbb R_{\geq 0}\},  
\end{align}
where $\mathcal B^p_-=\mathcal P_p-\xi$ and $\mathcal B^p_+=\mathcal P_p+\xi$ are the lower and upper boundaries respectively, with an offset $\xi \geq 0$.
%
%
%
%
\subsection{Frequencies Prediction}
\label{subsec:EPSDPStacker}
%
%
Due to the fact, that variable lift position and load lead to time-varying vibrational behavior and thus time-varying resonance frequencies, parameter variation $\varrho$, given in Section \ref{sec:NaturalFrequencies} are considered in \eqref{eq:ODEMS2}.  
Let $\breve{q}=[a_1, a_2]^T \in \breve{Q}$ be a prediction vector of states $a_1$ and $a_2$. From \eqref{eq:AxialDeflection} its obvious that these states, together with the spatial variable $y$ describe the vibration of the beam in terms of axial deflection. Based on the state predictions in $\breve{Q} \in \mathbb R^{2\times N}$, it is possible to predict the vibrational behavior and the associated frequencies. 
%
%

%
%
\section{Results: Stacker Crane}
\label{sec:result}
%
%
%
The optimal control problem (OCP), to move from an initial state to some terminal state, is stated as a time-OCP and solved in terms of NMPC in closed-loop control fashion. The vibrational behavior is considered explicitly to avoid undesired vibrational frequencies. Initial $x_0$ and terminal state $x_f$ are formulated as follows. Both, the carriage and lift are required to move from $m_c=\unit[-0.5]{m}$ to $\unit[0]{m}$ and from $y_l=\unit[0]{m}$ to $\unit[1]{m}$, respectively, under an initial deviation of the tip of $\omega_t=\unit[0.5]{m}$. 
Moreover, the STC is subject to constraints, such as on the carriage position with $\|x_c\|_{\infty}\leq \unit[0.5]{m}$, lift position $\unit[0]{m}\leq |y_l|\leq \unit[2]{m}$ and control input $\|F\|_{\infty}\leq \unit[30]{N}$. Both the carriage and the lift are actuated in terms of $F_1$ and $F_2$ respectively. In our simulation setup, both problems $\mathbb {P}_N(x_0, .)$ and $\mathbb {P}^Q_N(x_0, .)$, from Section \ref{sec:QuasiSoftMPC} and \ref{sec:SoftMPC}, are solved. For $\mathbb {P}_N(x_0, .)$, the weighting matrices of the stage and terminal costs of the objective function are chosen as $S=P=2\cdot 10^7 \cdot \I_2$ and $F=10^5$, with $s \in \RR^{2}$. Subsequently, the remaining weighting matrices from $\mathbb {P}^Q_N(x_0, .)$ are chosen as, $Q=\text{blkdiag}\{2000,10\} \otimes \I_4$, and $R= \I_2$. $x_s=x_f$ and $u_s=0_{2\times 1}$. 
For \eqref{eq:PolyApprox} the offset is considered as $\xi=10$. Following parameter are selected with corresponding SI units $m_c=\unit[2.888]{}, m_t=\unit[0.5]{}, m_l=\unit[1]{}, L=\unit[2]{}, A=\unit[3.2\cdot 10^{-4}]{}, \rho=\unit[2700]{}, EI=\unit[119.4]{}$. For NMPC scheme, a sampling time of $T_s=\unit[0.03]{sec}$ is considered and the optimization horizon is equal to 17. For the frequency prediction, and due to the sinusoidal nature of beam vibrations a sinusoidal fitting function is used before applying PSD, as in Section \ref{subsec:EPSDP} for accuracy improvement. Additionally, the corresponding window for PSD has a Gaussian function. 
%
%
%
%
%
%
\begin{figure}[htbp]
\begin{center}
\includegraphics[width=0.5\textwidth]{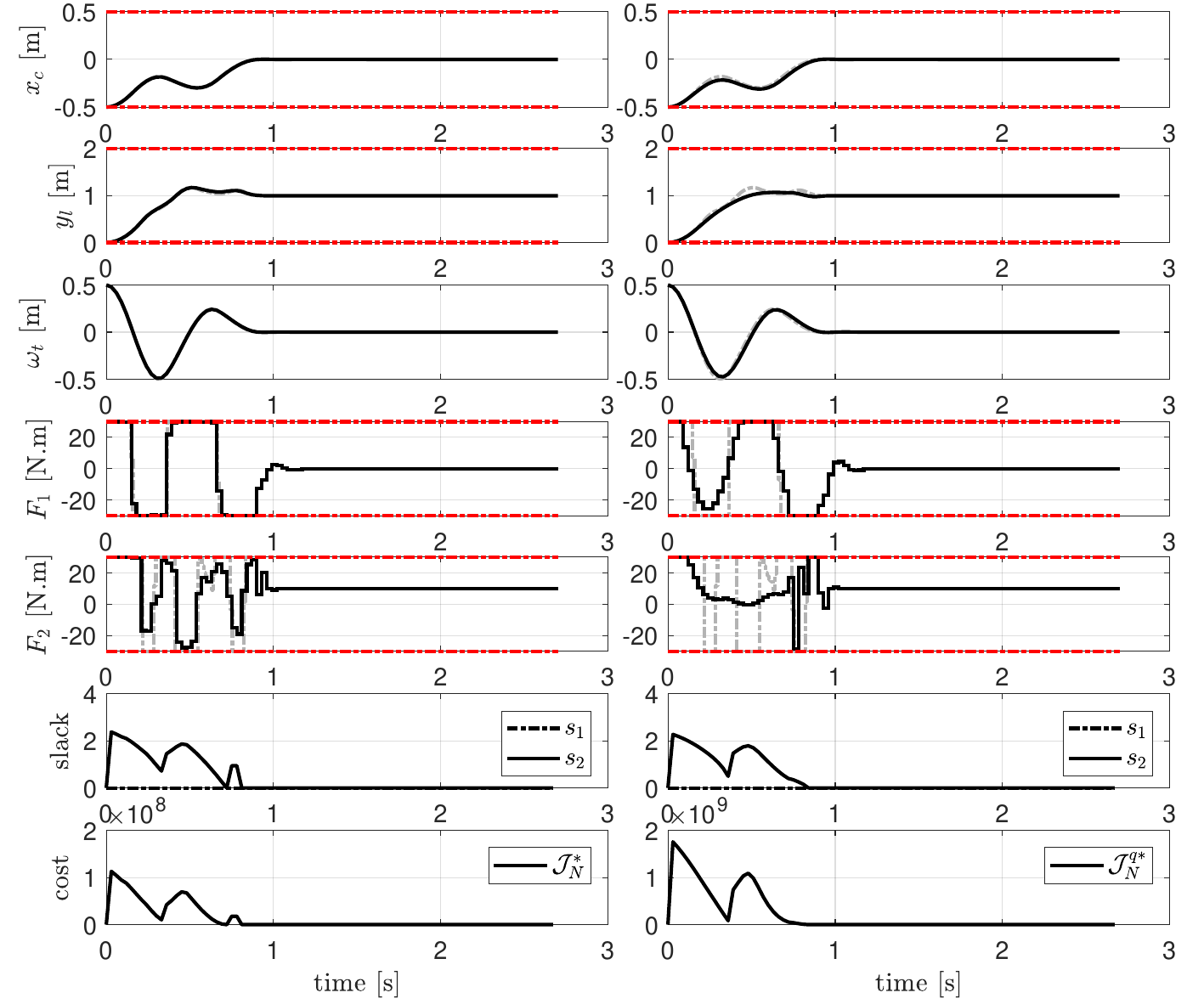}
\vspace{-2mm}
\caption{active vibration damping control for STC with soft NMPC due to $\mathbb {P}_N(x_0, .)$ (left) and $\mathbb {P}^Q_N(x_0, .)$ (right)} 
\label{fig:NMPC}
\end{center}
\end{figure}
%
%
%
%
%
The simulation results of the soft-constrained NMPC is showed in Fig. \ref{fig:NMPC} for both problems $\mathbb {P}_N(x_0, .)$ and $\mathbb {P}^Q_N(x_0, .)$. The closed-loop results with the corresponding states, control inputs, slack variables and costs are plotted. 
Note that, the purpose of this simulation is to demonstrate the performance of the soft-constrained time-optimal NMPC as an approach on both time-optimal problems and not to draw a comparison between both problems. Due to the different objectives and weightings, no fair comparison can be established between both problems. 
However, it is clear, that $\mathbb {P}_N(x_0, .)$ target mainly the time optimality, while $\mathbb {P}^Q_N(x_0, .)$ target time optimality while penalizing the state and control errors. For both problems, terminal sets $\X_f$ are designed as a control positive invariant set using the solution of the discrete Riccati equation of the linearized system dynamics around the terminal state $x_f$ to obtain recursive feasibility. The soft-constrained formulation enlarges the feasible set compared to an only hard constrained problem, which is under these given conditions not feasible at all. According to Fig. \ref{fig:NMPC}, during the position change of the lift $y_l$, different resonance frequencies were predicted. For both problems $\mathbb {P}_N(x_0, .)$ and $\mathbb {P}^Q_N(x_0, .)$ the slack variables are active every time, when the predicted frequencies violate the constraints. The corresponding violation contributes, together with the running and terminal costs, crucially into the cost. The curve of slack variables shows a decreasing behavior with some slumps, which are related to inaccuracy in frequency prediction when open-loop trajectories change the direction and it doesn't oscillate enough. The curves of the objectives are associated to the slack penalization, which show similar behaviors. The time spent to reach this terminal state is the resulting transition time $T$. After that, the controller in the invariant set is active. Control inputs ($F_1, F_2$) show the resulting forces on both lift and cart. In the invariant set, the corresponding controller holds the lift in its position by compensating the gravity.  
To give an impression of the extent of our approach, a point-to-point time-optimal simulation is conducted additionally and the corresponding states and inputs (dashed gray) are plotted on both $\mathbb {P}_N(x_0, .)$ and $\mathbb {P}^Q_N(x_0, .)$. The corresponding final states $x_f$ are reached after an optimal  transition time of $T=\unit[0.9133]{sec}$, which is the shortest compared to $\mathbb {P}_N(x_0, .), T=\unit[0.9160]{sec}$ and $\mathbb {P}^Q_N(x_0, .)$. Both soft-constrained NMPC trade a longer oscillation for less frequency excitations. As a result, the control inputs mitigate slightly the usual bang-bang behavior, especially a the corners between accelerations or decelerations.
The different of objectives in $\mathbb {P}_N(x_0, .)$ and $\mathbb {P}^Q_N(x_0, .)$ is visible in the curve propagation of the cost, while in $\mathbb {P}_N(x_0, .)$ only the slack and time propagation is apparent, the cost in $\mathbb {P}^Q_N(x_0, .)$ contains additionally the state and control errors, which decay smoothly.
For the states $x_c$, $y_l$ and both $a_1, a_2$ in forms of $\omega_t$, a good convergence in terms of closed-loop performance is shown. Starting with a big initial deviation of the beam at the tip, the NMPC is able to damp this vibration fast. Solutions obtained are feasible for both, hard (red dashed) and soft constraints. The computational effort is almost similar to a classical NMPC. In summary, by changing the inequality constraints \eqref{eq:NLPEquation4} and \eqref{eq:NLPEquation5} to include an arbitrary frequency, it can be shown, that a selective frequency damping is possible using this approach. For problem $\mathbb {P}_N(x_0, .)$ additional soft-constraint on the movement frequency of $y_l$ can be introduced to minimize such behavior. Note that, the frequency prediction is conducted based on the open-loop state trajectories, which have finite horizon and marginally different from the closed-loop state trajectories.   
%
%
%
%
%
%
%
%
%
%
%
\section{Additional case study: Fixed Boundary Frequencies}
\label{sec:result2}
%
%
In this example a two degree-of-Freedom spring-mass system with fixed resonance frequencies is considered. The stiffness and mass elements are assigned as, $k_1=k_2=k_3=\unit[1000]{N/m}$ and $m_1=m_2=
\unit[1]{kg}$. The equations of motion of the system can be written as,
\begin{align}
\Me{m_1& 0 \\ 0 & m_2} \Me{\ddot{x}_1 \\ \ddot{x}_2} + 
\Me{k_1+k_2& -k_2 \\ -k_2 & k_2+k_3} \Me{x_1 \\ x_2} = \Me{F_1 \\ F_2} 
\label{eq:2-DOF}  
\end{align}
The eigenvalue solution of \eqref{eq:2-DOF} produces the eigenvalues and eigenvectors, which show the natural frequencies and mode shapes of the system. The natural frequencies are given as $w_1 = \unit[5.033]{Hz}$ and $w_2 = \unit[8.717]{Hz}$. As it is stated in Section \ref{sec:ProblemFormulation}, the purpose of the controller is to smoothly steer the system from some initial state $x_0$ to a neighborhood of a terminal set within a minimal transition time T and with the minimal possible vibrations around the resonance frequencies. It is worth to mention, that the system is actually a highly undamped harmonic oscillator. Accordingly, two soft inequality constraints ($\mathcal B^p_-, \mathcal B^p_+$) of undesired frequencies in \eqref{eq:NLPEquation4} and \eqref{eq:NLPEquation5} are designed with following low boundaries $\mathcal B^p_{-,1} = w_1$ and $\mathcal B^p_{-,2} = w_2$. Both are time-invariant frequencies, which differ from the previous case study.
For initial states $x_1=\unit[1]{m}, x_2=\unit[-1]{m}$ and under constraints on $\|x_{1,2}\|_{\infty}\leq \unit[1]{m}$ and input $\|F\|_{\infty}\leq \unit[200]{N}$, two time-optimal NMPC problems are solved. The first one is a classical time-OCP problem formulated in NMPC scheme and the second one is the soft-constrained quasi time-optimal NMPC proposed in \ref{sec:SoftMPC}. For both NMPC schemes, a sampling time of $T_s=\unit[0.02]{sec}$ and horizon of 17 are considered. The weighting matrices of the stage and terminal costs of the objective function are chosen as $S=P=2\cdot 10^7 \cdot \I_2$. The simulation results are showed in Fig. \ref{fig:NMPC} for both problems. Plots for the time-optimal NMPC are showed in solid gray and dotted black corresponds to the quasi time-optimal NMPC. It is obvious that time-optimal NMPC converges faster, however with very small residual vibrations, which induce a bang-bang control. In contrast, the soft-constrained quasi time-optimal NMPC shows a mitigated bang-bang behavior and cancel the residual vibrations. Note that, the results are closed-loop results, while cost plots are open-loop costs.  
%
%
%
%
%
%
\begin{figure}[htbp]
\begin{center}
\includegraphics[width=0.5\textwidth]{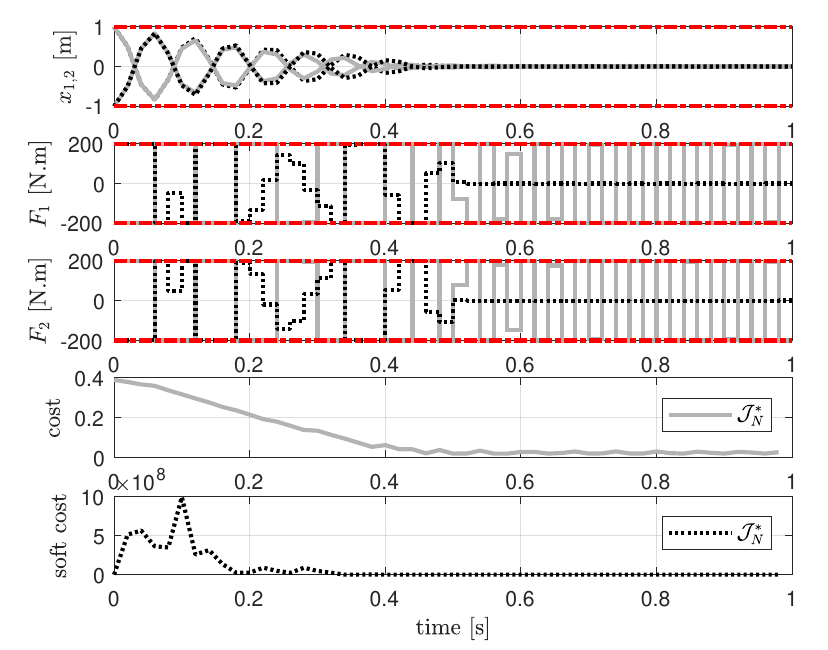}
\vspace{-2mm}
\caption{active vibration damping control with time-optimal (solid gray) and soft-constrained time-optimal (dotted) NMPC} 
\label{fig:NMPC2}
\end{center}
\end{figure}
%
%
%
%
%
%
\section{CONCLUSION} 
\label{sec:conc}
In this paper a new time-optimal soft-constrained nonlinear model predictive control scheme is presented and applied for the problem of active vibration damping control of flexible stacker cranes. This scheme deals with problems arising by classical time-optimal approach and the resulting bang-bang control input. In particular, the problem of exciting undesired vibrations is addressed in this scheme. The idea of the proposed scheme, is to define inequality constraints for undesired frequencies. By future prediction and constraints relaxation any violation is penalized in the cost. By containing any arbitrary frequency also for the state trajectory, this approach can be considered for a selective frequency damping or state trajectory penalization. Extension to guarantee the recursive feasibility and hardware implementation is in sight. 
%
%
%
\bibliography{Chapter/LiteratureBib}             
%
%
%
%
%
%
%
\end{document}